\documentclass[aps,pre,reprint,showpacs]{revtex4-1}
\usepackage{amsmath,graphicx}
\begin{document}

\setlength{\tabcolsep}{4pt}

\title{Long polymers near wedges and cones}
\author{Yosi Hammer}
\email{Email: hammeryosi@gmail.com}
\author{Yacov Kantor}
\affiliation{Raymond and Beverly Sackler School of Physics and
Astronomy, Tel Aviv University, Tel Aviv 69978, Israel}

\begin{abstract}
We perform a Monte Carlo study of $N$-step self-avoiding walks, attached to the corner of an impenetrable wedge in two dimensions ($d=2$), or the tip of an impenetrable cone in $d=3$, of sizes ranging up to $N=10^6$ steps. We find that the critical exponent $\gamma_{\alpha}$, which determines the dependence of the number of available conformations on $N$ for a cone/wedge with opening angle $\alpha$, is in good agreement with the theory for $d=2$. We study the end-point distribution of the walks in the allowed space and find similarities to the known behavior of random walks (ideal polymers) in the same geometry. For example the ratio between the mean square end-to-end distances of a polymer near the cone/wedge and a polymer in free space depends linearly on  $\gamma_{\alpha}$, as is known for ideal polymers. We show that the end-point distribution of polymers attached to a wedge does not separate into a product of angular and radial functions, as it does for ideal polymers in the same geometry. The angular dependence of the end-position of polymers near the wedge differs from theoretical predictions.
\end{abstract}

\date{\today}


\maketitle

\section{Introduction}

The statistics of the polymer conformations \cite{Gennes1979,Eisenriegler1993} in the presence of confining geometry has been the subject of extensive study for many years. Of particular interest is the case of long polymers, which is  closely related to the critical phenomena of magnetic systems \cite{Gennes1979,Li1995,Cardy1993,Eisenriegler1993}. Some properties of the polymers are \emph{universal}, i.e., close to the critical point (when the number of monomers in the polymer $N\rightarrow\infty$), they are independent of most of the details in the system. 
An important model for polymers is a lattice walk, which captures all the universal features \cite{Gennes1979}. Random walks (RWs) are used to model \emph{ideal} polymers, where different monomers are allowed to inhabit the same volume in space. Self-avoiding walks (SAWs) are used to model polymers in good solvent, where steric interaction between monomers exists. 
For large $N$, the number of configurations of a free $N$-step lattice walk starting from the origin, \cite{Madras1993}
\begin{equation} \label{eq:gammaDef}
\mathcal{N}_{\text{f}}\propto N^{\gamma_{\text{f}}-1}\mu^N,
\end{equation}
where $\gamma_{\text{f}}$ is a universal constant (exponent), and $\mu$ is the coordination number of the particular lattice for RWs or effective coordination number for SAWs. 

\begin{figure} 
\includegraphics[width=8cm]{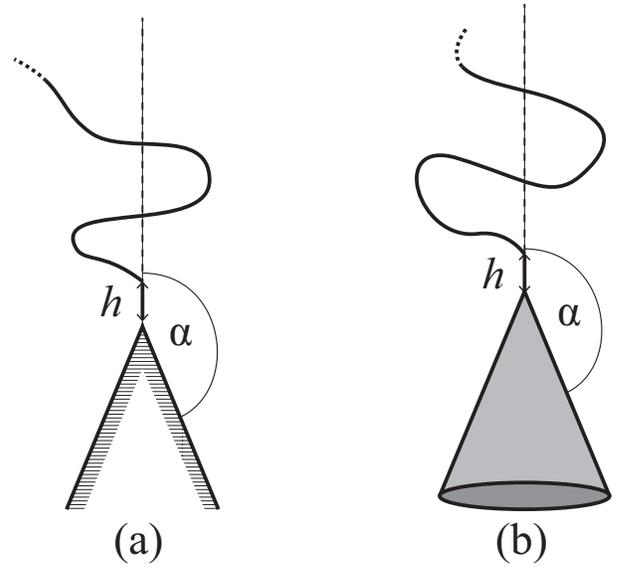}
\caption{Geometries considered in this paper. (a) A two-dimensional wedge with opening angle $\alpha$. (b) A three-dimensional cone with opening angle $\alpha$.}
\label{fig:geometry}
\end{figure}

In this work we study the behavior of long polymers attached to an excluded infinite cone in dimension $d=3$ or wedge in $d=2$ (see Fig.~\ref{fig:geometry}). These objects are \emph{scale-free} \cite{Maghrebi2011,Maghrebi2012,Hammer2014,Alfasi2015,Colby1987,Gaunt1990}, i.e., their shape is invariant with respect to a re-scaling  by an arbitrary factor $\lambda$, i.e., $\vec{r}\rightarrow\lambda\vec{r}$. Generally, such scale transformations change the position of the surface. Here we study polymers attached to the tip of the cone/wedge, which is placed at the origin, so that the position of the surface and the starting position of the polymer will not change under a scale transformation. The number of polymer conformations near the cone/wedge  \cite{Cardy1984,Guttmann1984,Hammersley1985,Maghrebi2011,Maghrebi2012,Hammer2014}
\begin{equation} \label{eq:gammaAlphaDef}
\mathcal{N}_{\alpha}\propto N^{\gamma_{\alpha}-1}\mu^N,
\end{equation}
where $\gamma_{\alpha}$ depends on the opening angle $\alpha$ of the cone/wedge (but does not depend on any microscopic details of the polymer), while $\mu$ is the same as in Eq.~\eqref{eq:gammaDef} \cite{Hammersley1985}. For ideal polymers, $\gamma_{\alpha}$ can be found analytically by solving the diffusion equation in the relevant geometry \cite{Ben-Naim2010,Maghrebi2011,Maghrebi2012,Hammer2014,Alfasi2015}. For SAWs near a wedge ($d=2$), $\gamma_{\alpha}$ was calculated \cite{Cardy1984} using  the analogy between magnetic and polymer systems near the critical point, along with conformal invariance \cite{WegnerF1976,Cardy1993,Li1995}. In Sec.~\ref{sec:deltaGamma} we measure the difference $\Delta\gamma_{\alpha}=\gamma_{\text{f}}-\gamma_{\alpha}$ using exceptionally long walks ($N$ as high as $10^6$), and compare our results with theoretical predictions.

\begin{figure}
\includegraphics{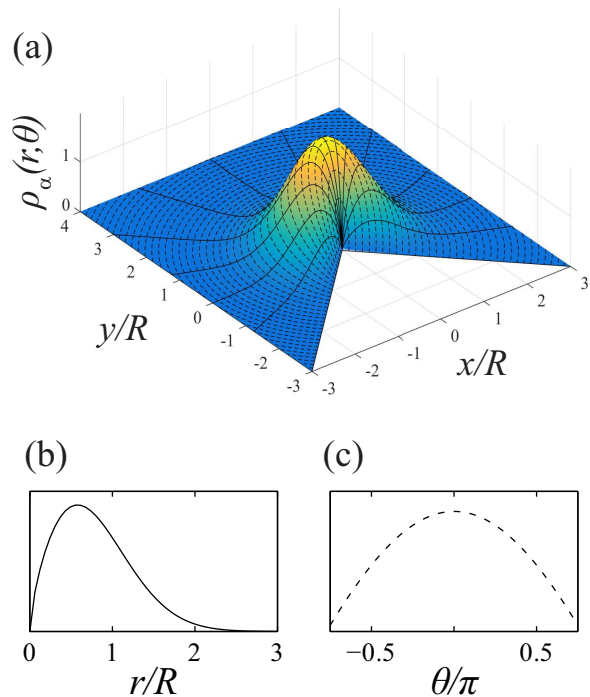}
\caption{(Color online) End-point distribution of a long two-dimensional ideal polymer (arbitrary units) attached to the corner of an excluded wedge with opening angle $\alpha=3\pi/4$. The distance from the corner is scaled with the root mean square end-to-end distance of a polymer in free space $R$. The distribution is composed of a product of a radial and angular functions. The solid lines in (a) represent curves with constant $\theta$, which all have the same form depicted in (b). Similarly curves with constant $r$ are represented by dashed lines in (a) and their form is shown in (c).}
\label{fig:idealEndPoint}
\end{figure}

In Sec.~\ref{sec:endPointConfined} we study the end-point distribution $\rho_{\alpha}(\vec{r})$ of a SAW in the wedge/cone, and the mean square end-to-end distance $R_{\alpha}^2$. It is known that $R_{\alpha}^2\propto R^2\propto N^{2\nu}$ \cite{Gennes1979}, where $R^2$ is the mean square end-to-end distance of the polymer in free space, and the exponent $\nu$ is independent of $\alpha$. 
For long ideal polymers attached to wedges/cones, $\rho_{\alpha}(\vec{r})$ is a product of radial and angular functions \cite{Hammer2014,Alfasi2015}:
\begin{equation} \label{eq:idealDistribution}
\rho_{\alpha}(\vec{r})\propto r^{\Delta\gamma_{\alpha}/\nu}\exp\left[-\frac{d}{2}\left(\frac{r}{R}\right)^2\right]\Theta_{\alpha}(\hat{r}),
\end{equation}
where the function $\Theta_{\alpha}$, which is known for a cone and a wedge, depends only on the direction of $\vec{r}$ ($\hat{r}$ is a unit vector). Note that we assume the distance from the anchor point of the polymer to the tip of the wedge/cone $h\ll R$ and therefore appears in $\rho_{\alpha}$ only as a prefactor which we neglected in Eq.~\eqref{eq:idealDistribution}. Fig.~\ref{fig:idealEndPoint} depicts $\rho_{\alpha}$ for an ideal polymer near a wedge with $\alpha=3\pi/4$. The solid lines in Fig.~\ref{fig:idealEndPoint}a represent curves with constant $\theta$. They all have the same form shown in Fig.~\ref{fig:idealEndPoint}b. Similarly, the dashed lines in Fig.~\ref{fig:idealEndPoint}a represent curves with constant radius, which have the form shown in Fig.~\ref{fig:idealEndPoint}c. Note that for ideal polymers, $\Delta\gamma_{\alpha}$ determines the number of polymer conformations, and also determines the small $r$ behavior of $\rho_{\alpha}$. Moreover, Eq.~\eqref{eq:idealDistribution} leads to 
\begin{equation} \label{eq:universalRatio}
R_{\alpha}^2/R^2=\frac{\Delta\gamma_{\alpha}}{d\nu}+1.
\end{equation}
For SAWs, however, not much is known about $\rho_{\alpha}$, and that is the focus of our study. We wish to see which of the properties of ideal polymers in scale free spaces carries over to polymers in good solvent. 

Monte Carlo simulations face a challenge to generate large ensembles. The pivot algorithm \cite{Lal1969,Madras1988,Kennedy2002,Clisby2010,Clisby2010b} is a dynamic method which generates SAWs with fixed $N$ and free end-points. In each time step a random site along the walk is used as a pivot point for a random symmetry action on the lattice (e.g., rotation or reflection) to be applied to the part of the walk subsequent to the pivot point. The resulting walk is accepted if it is self-avoiding; otherwise, it is rejected and the old walk is sampled again. The pivot algorithm is most efficient when studying large scale properties of the polymers such as $R^2$ \cite{Madras1988}. Still, the bottleneck in the algorithm was always the self-avoidance tests, which required $O(N^x)$ operations ($x\sim1$). 

Recently, Clisby \cite{Clisby2010,Clisby2010b} introduced a new data structure called a SAW tree that allows for a faster implementation of the pivot algorithm. An $N$-site SAW is represented by a binary tree \cite{Cormen2009}. The $N$ leafs of the tree are individual sites of the walk, while each of the $N-1$ internal nodes represents a section of the walk. Each internal node contains aggregate information about the section of the walk that it represents, such as the bounding box of the section, which is a convex shape that completely contains it. Each node also contains a symmetry operation. The walk corresponding to a SAW tree is constructed in a recursive procedure starting from the root and moving down the tree. At each step, the symmetry operation stored in the node is performed on the right child, and then the left and right children are concatenated. Clisby also defined a set of rotation operations \cite{Cormen2009} that can change the structure of the tree while not changing the SAW it represents. A pivot move can be implemented by selecting a node and changing the symmetry operation it contains, after using rotations to bring it to the root of the tree. In order to check for intersections between sections of the walk, one can recursively check for intersections between bounding boxes of left and right children in the tree. Clisby showed that on average, a remarkably low number of intersection tests is needed, and a test for self-intersection can be done in at most O($\log N$) time. As mentioned in \cite{Clisby2010b}, the SAW tree can be used to perform fast intersection tests with surfaces. This is done by recursively checking for intersections between the bounding boxes of left and right trees and a boundary surface. In this work, we implemented such intersection tests for wedges and cones which also take no longer than O($\log N$) time to perform. Note that while the self-intersection tests are used to accept or reject pivot moves, in our simulation, the intersection tests of walks and surfaces are used only to decide whether to include the walks in the statistics of a particular geometry (see Sec.~\ref{sec:deltaGamma}).

The improved implementation of the pivot algorithm enables the study of SAWs in confined spaces of sizes that were not accessible to simulations in the past. In Sec.~\ref{sec:deltaGamma} we explain the simulation method, review the measurement of $\Delta\gamma_{\alpha}$, and compare with available theory. In Sec.~\ref{sec:EndPointFree} we discuss the end-point distribution of a SAW in free space  $\rho_{\text{f}}(\vec{r})$ as a preparation for the following discussion of the confined distribution. In Sec.~\ref{sec:endPointConfined} we discuss the end-point distributions for SAWs starting near a wedge/cone and find disagreement with the predicted angular dependence of the end-point position.  Sec.~\ref{sec:ratio} is devoted to the universal ratio $R_{\alpha}^2/R^2$,  where we find that the RW behavior can be carried over to the case of SAWs quite well. Final conclusions and remarks are found in Sec.~\ref{sec:conclusions}.

\section{Measurement of $\Delta\gamma_{\alpha}$}
 \label{sec:deltaGamma}
In order to study the behavior of the exponent $\gamma_{\alpha}$, we measured the probability $p_{\alpha}(N)$ of successfully moving a lattice walk with $N$ steps that was generated in free space to the vicinity of an excluded wedge/cone with opening angle $\alpha$ (without changing its shape). This probability is equal to the fraction of walks that, starting near the tip of the cone/wedge, do not intersect with the boundary. From Eqs.~\eqref{eq:gammaDef} and \eqref{eq:gammaAlphaDef}, 
\begin{equation} \label{eq:pn}
p_{\alpha}(N)=\frac{\mathcal{N}_{\alpha}}{\mathcal{N}_{\text{f}}} \propto  N^{-\Delta\gamma_{\alpha}}.
\end{equation}

We used the pivot algorithm to generate at least $2\times10^6$ SAWs in free space with $N=10^5$ to $10^6$, where sequential samples were separated by $10^3$ pivot attempts (Other methods to manage large highly correlated data could also have been used, such as time batching \cite{Geyer2011}). Each simulation began with an initialization of the SAW in order to avoid any systematical errors due to its initial configuration. Madras and Sokal \cite{Madras1988} estimated that it is sufficient to discard the first $20N/f$ time steps in the simulation, where $f$ is the acceptance rate of the pivot moves. Given that the lowest acceptance rate in our simulation was $f\sim 0.07$ (for SAWs on the square lattice with $N=10^6$) we found it sufficient to discard the first $500N$ time steps in all configurations studied in this work. We selected the pivot points uniformly from the sites along the walk. Reduction of the simulation time or an increase in accuracy can be achieved by using non-uniform selection of the pivot sites along the walk, thus reducing the autocorrelation time for the samples in the simulation \cite{Clisby2015}. 
 
 In every sample the walk was tested for intersection with a wedge/cone with various opening angles $\alpha$, placed so that the starting position of the walk is ten lattice units from the tip. We measured the probability $p_{\alpha}(N)$ that the walk will not intersect with the boundary and the distribution of the end-point of the walk (used in Sec.~\ref{sec:endPointConfined}). Note that the pivot attempts were performed in \emph{free} space and, when studying the end-point distribution of a SAW in a particular wedge/cone, only those samples in which the walk did not intersect with the surface were considered. Since the pivot algorithm in free space is ergodic, all samples in the generated ensemble are taken into account with the same statistical weight. The same is true for a subset of this ensemble where the walks do not intersect a certain boundary. 
 
In order to verify the validity of our method, we repeated the simulation for RWs, this time accepting any pivot attempt, without checking for self-avoidance, and compared our data with known theoretical results. Note that Eqs.~\eqref{eq:gammaDef} and \eqref{eq:gammaAlphaDef} are valid in the asymptotic limit $N\rightarrow\infty$ and, in principle, we should consider finite size corrections (For example $\mathcal{N}_{\text{f}}\propto N^{\gamma_{\text{f}}-1}\mu^N\left(1+O(N^{-\Delta})\right)$, where $\Delta\approx0.52$ \cite{Clisby2013} in $d=3$). However, since the smallest $N$ in this simulation is $10^5$, we find that finite size corrections are negligible compared to the errors due to the statistical scattering of the data.

\begin{figure} 
\includegraphics{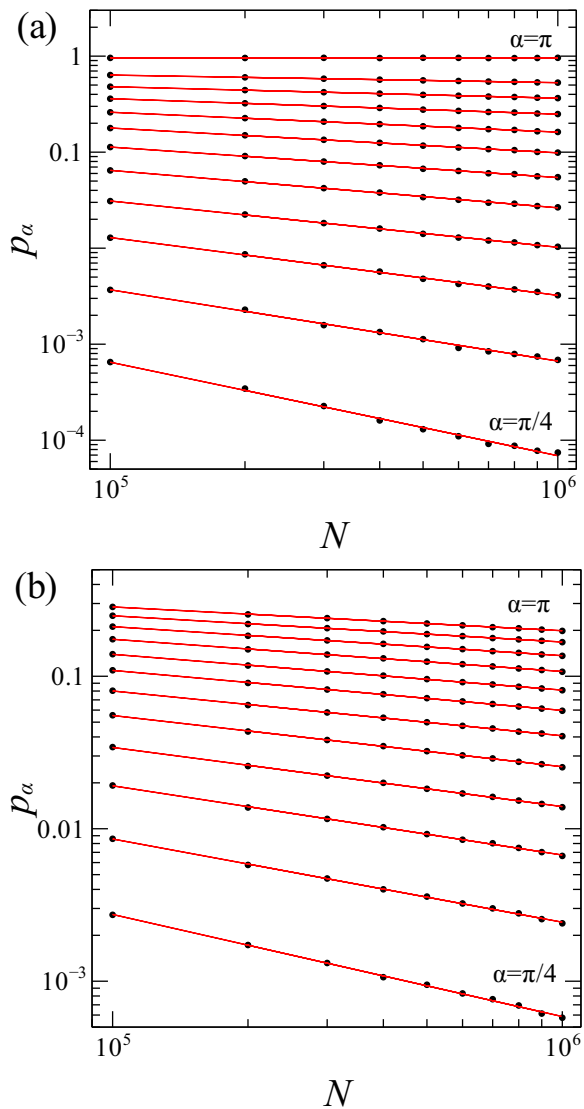}
\caption{(Color online) Success probabilities $p_{\alpha}$ for the confinement of SAWs of size $N$ in a three-dimensional cone (a) or a two-dimensional wedge (b) with different opening angles $\alpha=j \pi/16$, $j=4,5,...,16$. The continuous lines are linear fits.}
\label{fig:SuProb}
\end{figure}

\begin{figure}
\includegraphics{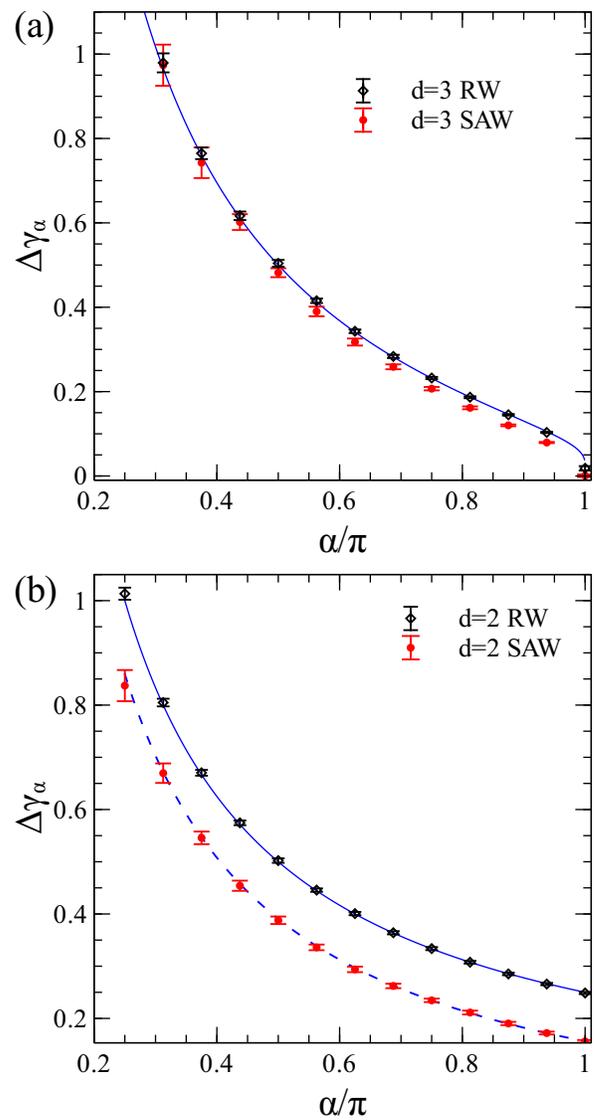}
\caption{(Color online) Difference in the critical exponents $\gamma-\gamma_{\alpha}$ as a function of the opening angle of the cone in dimension $d=3$ (a) or wedge in $d=2$ (b), for RWs (black diamonds) and SAWs (red circles). The dashed line represents the prediction by Cardy and Redner \cite{Cardy1984} for two-dimensional SAWs, while the solid lines represent known values for RWs \cite{Maghrebi2011,Maghrebi2012,Hammer2014,Alfasi2015}.}
\label{fig:gamma}
\end{figure}

In Fig.~\ref{fig:SuProb} we present, in logarithmic scales, the success probabilities as a function of the size of the walk, for SAWs in $d=2$ and $d=3$.
The linear dependence is clearly observed. The absolute values of the slopes of the linear fits shown in Fig.~\ref{fig:SuProb} represent the difference in the critical exponents, $\Delta\gamma_{\alpha}$, and they are presented in Fig.~\ref{fig:gamma} (red circles).
\begin{table}
\begin{tabular}{ c  c }
 $\alpha/\pi$  & $\Delta\gamma_{\alpha}$ \\ \hline \hline
5/16 & 0.973 $\pm$ 0.049\\ \hline
3/8 & 0.743 $\pm$ 0.036\\ \hline
7/16 & 0.602 $\pm$ 0.019 \\ \hline
1/2 & 0.482 $\pm$ 0.011 \\ \hline
9/16 & 0.39 $\pm$ 0.012 \\ \hline
5/8 & 0.3175 $\pm$ 0.0083 \\ \hline
11/16 & 0.2588 $\pm$ 0.0058 \\ \hline
3/4 & 0.207 $\pm$ 0.004 \\ \hline
13/16 & 0.162 $\pm$ 0.0034 \\ \hline
7/8 & 0.12 $\pm$ 0.002\\ \hline
15/16 & 0.0793 $\pm$ 0.0014 \\ \hline
1 & 0.0005 $\pm$ 0.0031 \\
\end{tabular}
\caption{Difference in the critical exponents $\gamma-\gamma_{\alpha}$ measured for SAWs in $d=3$ confined to a cone with opening angle $\alpha$ (red circles in Fig. \ref{fig:gamma}a).}
\label{tab:deltaGamma}
\end{table}
A similar analysis for RWs is also shown in Fig.~\ref{fig:gamma} (black diamonds). The dashed line in Fig.~\ref{fig:gamma}b represents the prediction from conformal invariance \cite{Cardy1984}, while the solid lines in Figs.~\ref{fig:gamma}a and \ref{fig:gamma}b represent the known values of $\Delta\gamma_{\alpha}$ for RWs in two and three dimensions \cite{Maghrebi2011,Maghrebi2012,Hammer2014,Alfasi2015}. We find excellent agreement with all theoretical predictions. For SAWs in $d=3$, where there are no analytical estimates for $\Delta\gamma_{\alpha}$, we present our numerical estimates in Table \ref{tab:deltaGamma}. Recently Clisby \textit{et al.} \cite{Clisby2015} used the pivot algorithm to study SAWs near a flat plane ($\alpha=\pi/2$) and reached the very accurate estimate $\Delta\gamma_{\alpha}=0.479315(20)$, in agreement with our result $\Delta\gamma_{\alpha}=0.482\pm0.011$. Note that in \cite{Clisby2015} a specific geometry for the boundary was studied while in this work we study a range of excluded surfaces. 

It is interesting to observe the dependence of the critical exponents on the dimensionality and the presence of self-avoidance when the cone/wedge is reduced to a semi-infinite line (when $\alpha\rightarrow\pi$) \cite{Caracciolo1997,SCaraccioloandGFerraroandAPelissetto1991,Considine1989,Douglas1989}. In $d=2$, a semi-infinite line is a significant barrier to the walk and the change in the critical exponent $\Delta\gamma_{\pi}>0$. (In the language of renormalization group, one can say that in this case, the boundary constitutes a \emph{relevant} perturbation on the free space Hamiltonian \cite{Caracciolo1997,Douglas1989}). In $d=3$, the semi-infinite line is not a significant barrier and does not change the critical exponents. However, for ideal polymers in $d=3$, $\Delta\gamma_{\alpha}$ approaches zero like $1/\ln|\pi-\alpha|$. (In this case the semi-infinite line constitutes a \emph{marginal} perturbation). As can be seen in Fig.~\ref{fig:gamma}a (solid line), the approach to zero is almost vertical. Note that the values shown for $\alpha/\pi=1$ were measured for RWs and SAWs with an excluded semi-infinite line. For SAWs we measured $\Delta\gamma_{\pi}=0$, while for RWs, the discrete space results in a small error in the opening angle of the cone that is significant due to the diverging derivative of $\Delta\gamma_{\alpha}$ near $\alpha=\pi$, and we measured $\Delta\gamma_{\pi}=0.018$.

\section{End-point distribution of a free SAW} \label{sec:EndPointFree}

To set the stage for the study of $\rho_{\alpha}(\vec{r})$, we start with the end-point distribution for a SAW starting from the origin in \emph{free} space $\rho_{\text{f}}(\vec{r})$, normalized so that $\int\rho_{\text{f}}(\vec{r})d^dr=1$. The free space distribution was studied extensively over the years. In the limit $N\rightarrow\infty$, $\rho_{\text{f}}(\vec{r})$ depends on $\vec{r}$ only through the ratio $\vec{r}/R$ \cite{Gennes1979}, 
\begin{equation} \label{eq:freeDistScaling}
\rho_{\text{f}}(\vec{r}) = \frac{1}{R^d}f(r/R),
\end{equation}
where the prefactor $1/R^d$ is required for normalization. Several properties of the function $f(x)$ are known \cite{Domb1965,Domb1965b,McKenzie1971,McKenzie1973,DesCloizeaux1974}: For large values of $x$, 
\begin{equation} \label{eq:largeXdependence} 
f(x)\propto x^{t_{\text{f}}}\exp\left(-Dx^{\delta_{\text{f}}}\right),
\end{equation}
where 
\begin{equation} \label{eq:t_f}
t_{\text{f}}=(\nu d-\gamma_{\text{f}}+1-d/2)/(1-\nu)
\end{equation}
and 
\begin{equation} \label{eq:d_f}
\delta_{\text{f}}=1/(1-\nu).
\end{equation} 
A similar exponential cutoff might be expected to hold for a polymer near an excluded cone/wedge, since the conformations of sizes significantly exceeding $R$ are very rare. This effect is not expected to be modified significantly by the presence of an excluded boundary near one of the polymer ends.
In the limit $x\rightarrow0$, $f(x)$ describes the chance of a SAW to return to the origin. It is known that in this case
\begin{equation} \label{eq:smallXdependence}
f(x)\propto x^{g_{\text{f}}},
\end{equation}
where 
\begin{equation} \label{eq:g_f}
g_{\text{f}}=(\gamma_{\text{f}}-1)/\nu.
\end{equation} 
Note that these results carry over to ideal polymers, where, since $\gamma=1$ and $\nu=1/2$, the powers $t_{\text{f}}=g_{\text{f}}=0$, i.e. the distribution does not vanish for $\vec{r}\rightarrow0$. Also for ideal polymers $\delta_{\text{f}}=2$ and we recover the Gaussian distribution. Note that the Gaussian form at large $x$ for an ideal polymer attached to a wedge/cone (Eq.~\eqref{eq:idealDistribution}) is independent of the opening angle of the cone, as expected.
The power law dependence in Eq.~\eqref{eq:smallXdependence} is present when the system has no characteristic length scale, and thus we expect to find it again when the polymer is attached to a scale free surface like a cone or wedge. However, the dependence of the power $g_{\text{f}}$ on the the other exponents, given in Eq.~\eqref{eq:g_f} relies on the translation invariance of the system \cite{Gennes1979,DesCloizeaux1974} and is not expected to hold when the boundary is introduced. Indeed, from Eq.~\eqref{eq:idealDistribution} we see that for an ideal polymer, the power law dependence at small $r$ does not agree with Eq.~\eqref{eq:g_f}. 

\begin{figure}
\includegraphics{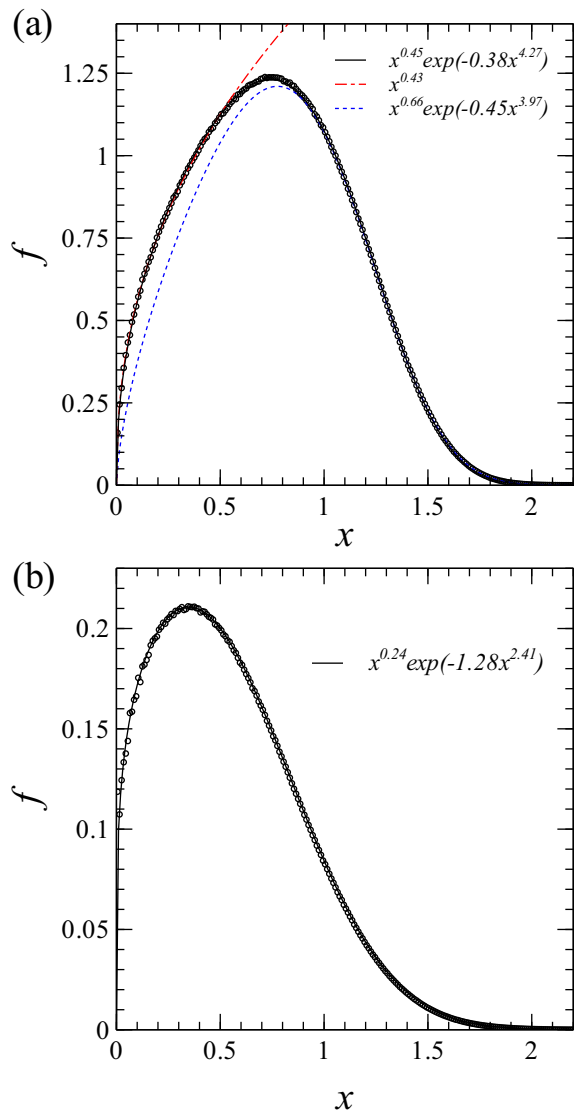}
\caption{(Color online) Scaled end-point distribution function $f(x)=R^d\rho_{\text{f}}(\vec{r})$ of a SAW in free space with (a) $10^5$ steps on a cubic lattice and (b) $5\times10^5$ steps on a square lattice, where $R^2$ is the mean square end-to-end distance of the walk and $x=r/R$. The solid lines are fits to the function in Eq.~\eqref{eq:phenoFitform}. The blue dashed line is a fit to a function of the same form performed only on the region $x>1.1$. The red dash-dot line is a fit to a simple power law (Eq.~\eqref{eq:powerFit}), performed only on the region $x<0.4$.}
\label{fig:freeRadialDist}
\end{figure}

When attempting to extract powers such as $t$, $g$ and $\delta$ from Monte Carlo simulations \cite{Dayantis1991,Everaers1995,Caracciolo2000}, we usually need to take into account two types of corrections to the analytical forms presented in Eqs.~\eqref{eq:freeDistScaling}, \eqref{eq:largeXdependence} and \eqref{eq:smallXdependence}. The first are corrections to scaling laws which result from the fact that the walks in the simulation are finite. For example $R^2\propto N^{\nu}(1+O(N^{-\Delta}))$, where $\Delta\approx1/2$ in $d=2,3$\cite{Caracciolo2000}. As we mentioned in Sec. \ref{sec:deltaGamma}, we find that such corrections are negligible here when compared to the statistical scattering of the data. The second are nonasymptotic corrections which result from the fact that Eqs.~\eqref{eq:largeXdependence} and \eqref{eq:smallXdependence} are expected to hold only in the regions $x\gg1$ and $x\ll1$ respectively. We attempted to estimate the effect of these correction by fitting different radial regions of the measured distributions separately. For small $x=r/R$ we used a simple power law of the form 
\begin{equation} \label{eq:powerFit}
f(x) = Ax^g,
\end{equation} 
with $A$ and $g$ the free parameters, and for the region of large $x$ we used the form 
\begin{equation} \label{eq:phenoFitform}
f(x)=Bx^t\exp\left(-Dx^{\delta}\right),
\end{equation}
with $B$, $t$, $D$, and $\delta$ as the free parameters. The cutoffs for the large and small $x$ regions where selected so that the best agreement with the theory for the exponents $g$, $t$ and $\delta$ is achieved. The same cutoffs were later used to analyze the behavior of SAWs attached the cones and wedges.  We also fitted the entire distribution to a function of the form given in Eq.~\eqref{eq:phenoFitform}, to see if it can be described by such a `phenomenological' function \cite{Dayantis1991,Everaers1995,Caracciolo2000}. Note that in principle, when using the entire distribution, the parameters in Eq.~\eqref{eq:phenoFitform} are not independent of each other due to normalization conditions. We do not take that into account directly in the fitting procedure but only indirectly by using properly normalized data. 
For SAWs in $d=3$ we find that attempting to fit different regions of the distribution separately does not alter or improve the results obtained by the phenomenological description and we only present the exponents derived from the phenomenological fit. This is not surprising since in $d=3$, $t_{\text{f}}\approx g_{\text{f}}\approx 0.26$ and both Eq.~\eqref{eq:largeXdependence} and Eq.~\eqref{eq:smallXdependence} can be satisfied by the same function. In $d=2$ we used $x>1.1$ for the large $x$ fit (blue dashed line in Fig.~\ref{fig:freeRadialDist}a) and $x<0.4$ for the small $x$ fit (red dash-dot line).
The exponents obtained from the fits to the free space end-point distributions are given in Tables \ref{tab:freeSpace2d} and \ref{tab:freeSpace3d}.
\begin{table}
\vspace{.4cm}
\begin{tabular}{ c   c  c  c }
 & $g$ & $t$ & $\delta$ \\ \hline \hline
theory & 0.458 &  0.625 & 4 \\ \hline
pheno. & - &  $0.4507\pm0.0016$ & $4.269\pm0.014$ \\ \hline
$x>1.1$ & - & $0.66\pm0.05$ & $3.97\pm0.07$ \\ \hline
$x<0.4$ & $0.436\pm0.003$ & - & - \\ \hline
\end{tabular}
\caption{Scaling exponents describing the end-point distribution of a free SAW in $d=2$ with $5\times10^5$ steps. The numerical values were obtained from the fits shown in Fig.~\ref{fig:freeRadialDist}a. The phenomenological results (pheno.) were obtained by fitting the entire distribution to the form given in Eq.~\eqref{eq:phenoFitform}. We also attempted to use only the region with $x>1.1$ to the same functional form while fitting a simple power law (Eq.~\eqref{eq:powerFit}) to the region $x<0.4$. The error estimates are derived from statistical scattering of the data and do not include systematical corrections mentioned in the text.}
\label{tab:freeSpace2d}
\end{table}
\begin{table}
\begin{tabular}{ c  c  c }
 & $t$ & $\delta$ \\ \hline \hline
free theory  &  0.260 & 2.427 \\ \hline
free numeric &  $0.2435\pm0.0019$ & $2.413\pm0.008$ \\ \hline
$\alpha=3\pi/8$ &  $1.408\pm0.047$ & $2.662\pm0.088$ \\ \hline
$\alpha=\pi/2$ &  $0.952\pm0.015$ & $2.615\pm0.036$ \\ \hline
$\alpha=5\pi/8$ &  $0.6883\pm0.0082$ & $2.560\pm0.023$ \\ \hline
$\alpha=3\pi/4$ &  $0.5278\pm0.0043$ & $2.505\pm0.013$ \\ \hline
$\alpha=7\pi/8$ &  $0.3923\pm0.0028$ & $2.4838\pm0.0094$ \\ \hline
$\alpha=\pi$ &  $0.2508\pm0.0021$ & $2.4078\pm0.0074$ \\ \hline
\end{tabular}
\caption{Scaling exponents describing the end-point distribution of SAWs in $d=3$ with $10^5$ steps in free space (Fig.~\ref{fig:freeRadialDist}b) or attached to the tip of an excluded cone with opening angle $\alpha$ (Fig.~\ref{fig:RadialDist}b). The numerical values were obtained from the fits shown in Fig.~\ref{fig:freeRadialDist}b and \ref{fig:RadialDist}b, were we used the function from Eq.~\eqref{eq:largeXdependence}. The error estimates are derived from statistical scattering of the data and do not include systematical corrections mentioned in the text.}
\label{tab:freeSpace3d}
\end{table}
We find good agreement with the theory apart from a slightly smaller value of $g_{\text{f}}$ and $t_{\text{f}}$ both in $d=2$ and $d=3$, probably due to the nonasymptotic corrections mentioned above. Note that the error estimates in the tables represent statistical scattering of the data but not systematical errors caused by non-asymptotic corrections for example. 

\section{End-point distribution for SAWs attached to wedges and cones} 
\label{sec:endPointConfined}
\subsection{Radial distribution} \label{subsec:radialDist}
\begin{figure}
\includegraphics{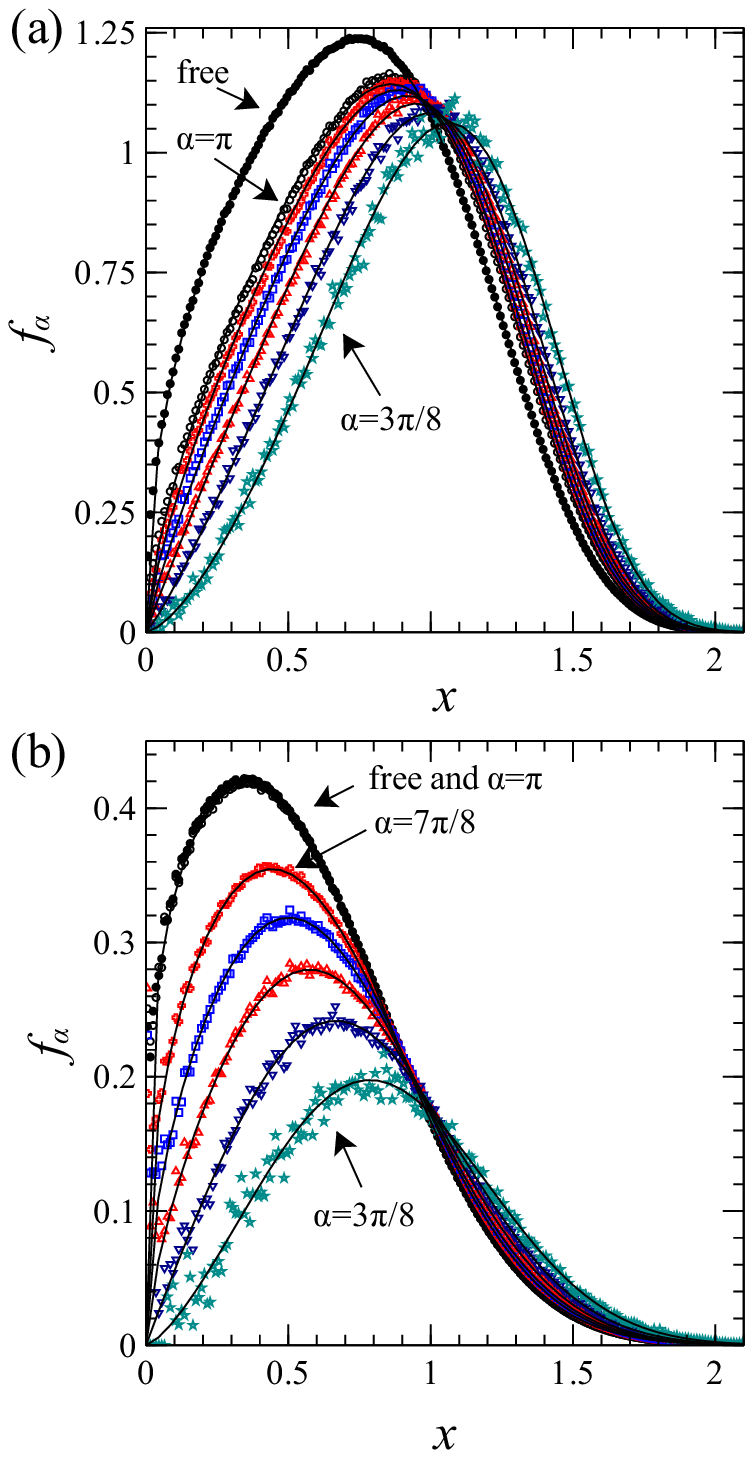}
\caption{(Color online) Scaled radial end-point distribution $f_{\alpha}(x) = R^d\rho_{\alpha}(r)$ of a SAW with (a) $10^5$ steps on a cubic lattice and (b) $5\times10^5$ steps on a square lattice starting near the tip of an excluded wedge/cone with opening angle $\alpha$, where $r$ is the distance from the tip of the wedge/cone, $R^2$ is the mean square end-to-end distance of the walk in free space and $x=r/R$. The distributions are plotted for $\alpha=i\pi/8$ with $i=3,4,..,8$. The solid lines are fits to the functional form in Eq.~\eqref{eq:phenoFitform}. We also plot the free space distributions $f(x)$ in $d=2$ and $2f(x)$ in $d=3$, where the factor two enters due to the azimuthal symmetry.}
\label{fig:RadialDist}
\end{figure}

In order to understand the behavior of SAW attached to wedges and cones, we attempt to separate the radial and the angular dependence of the end-point distribution, as can be done for RWs (Eq.~\eqref{eq:idealDistribution}). To study the radial behavior we define
\begin{align} \label{eq:radDistDef}
\rho_{\alpha}(r) \equiv \int_{-\alpha}^{\alpha} \rho_{\alpha}(r,\theta)d\theta \;\; (d=2), \\
\rho_{\alpha}(r) \equiv \int_{0}^{\alpha} \rho_{\alpha}(r,\theta)\sin\theta d\theta \;\; (d=3).\nonumber
\end{align}
We also define the scaled distribution $f_{\alpha}(x)=R^d\rho_{\alpha}(r)$, where $x=r/R$ as before. In Fig.~\ref{fig:RadialDist} we present the radial end-point distributions $f_{\alpha}(x)$ for SAWs starting from the tip of an excluded cone/wedge with opening angle $\alpha$. The exponents extracted from the fits are shown in Tables \ref{tab:freeSpace3d} and \ref{tab:wedge}.

\begin{table*}
\begin{tabular}{ c  c  c  c  c  c }
& \multicolumn{2}{ c }{pheno.} & \multicolumn{2}{ c }{$x>1.1$} & \multicolumn{1}{ c }{$x<0.4$}\\ \hline \hline
$\alpha$ & $t$ & $\delta$ & $t$ & $\delta$ & $g$ \\ \hline \hline
$3\pi/8$ & $1.500\pm0.022$ & $4.834\pm0.084$ & $1.64\pm0.42$ & $4.47\pm0.34$ & $1.54\pm0.11$ \\ \hline
$\pi/2$ & $1.1596\pm0.01$ & $4.794\pm0.048$ & $1.42\pm0.21$ & $4.24\pm0.16$ & $1.126\pm0.046$ \\ \hline
$5\pi/8$ & $0.9630\pm0.0063$ & $4.754\pm0.036$ & $1.54\pm0.16$ & $3.99\pm0.10$ & $0.916\pm0.019$\\ \hline
$3\pi/4$ & $0.8373\pm0.0056$ & $4.710\pm0.036$ & $1.51\pm0.13$ & $3.911\pm0.082$ & $0.782\pm0.014$\\ \hline
$7\pi/8$ & $0.7425\pm0.0051$ & $4.682\pm0.36$ & $1.26\pm0.10$ & $3.969\pm0.069$ & $0.6870\pm0.0092$\\ \hline
$\pi$ & $0.6736\pm0.0049$ & $4.659\pm0.037$ & $1.097\pm0.087$ & $4.00\pm0.059$ & $0.6166\pm0.0084$\\ \hline
\end{tabular}
\caption{Scaling exponents describing the end-point distribution of a SAW with $5\times10^5$ steps on a square lattice attached to the corner of an excluded wedge with opening angle $\alpha$. The values were obtained from the fits shown in Fig.~\ref{fig:RadialDist}a. The phenomenological results (pheno.) were obtained by fitting the entire distribution to the form given in Eq.~\eqref{eq:phenoFitform}. We also attempted to use only the region $x>1.1$ and fit to the same functional form while fitting a simple power law (Eq.~\eqref{eq:powerFit}) to the region $x<0.4$. The error estimates are derived from statistical scattering of the data and do not include systematical corrections mentioned in the text.}
\label{tab:wedge}
\end{table*}

Both in $d=2$ and $d=3$, $f_{\alpha}(x)$ can be closely approximated by a single function of the form given in Eq.~\eqref{eq:phenoFitform}. The power  $g$ at small radii increases dramatically when the opening angle is decreased (from 0.25 to 1.4 in $d=3$ and from 0.67 to 1.5 in $d=2$), while the exponential decline at large $x$ remains roughly the same. In the case of a semi-infinite line ($\alpha=\pi$), we see again that for a SAW in $d=3$ the effect of the line is negligible, and $f_{\pi}(x)$ is indistinguishable from $2f(x)$ (Fig.~\ref{fig:RadialDist}b). Note that the factor two enters here due to the azimuthal symmetry). In $d=2$, the distribution differs significantly from the free space form (Fig.~\ref{fig:RadialDist}a). We see that the radial dependence of polymers in good solvent is qualitatively similar to that of ideal polymers. The opening angle of the cone/wedge affects the exponent $g$, whereas the exponent $\delta$, related to large stretching of the polymer, is roughly independent of the surface to which the polymer is attached. The small variations in $\delta$ with respect to the opening angle of the cone are most likely a result of our fitting procedure, where we impose simple functional forms on the end-point distribution.

\subsection{Angular distribution in the wedge} \label{subsec:EndPointWedge}
We now turn to the angular distribution of end-points in the wedge ($d=2$). In \cite{Cardy1984} the angular distribution for a two-dimensional walk was defined,
\begin{equation} \label{eq:angDistDef}
\rho_{\alpha}(\theta)\equiv\int_0^{\infty} \rho(r,\theta)rdr,
\end{equation}
and it was predicted that for SAWs starting from the corner of an excluded wedge, 
\begin{equation} \label{eq:CardyAngular}
\rho_{\alpha}(\theta)\propto \left[\cos\left(\frac{\theta\pi}{2\alpha}\right)\right]^{\frac{\Delta\gamma_{\pi/2}}{\nu}}.
\end{equation}
Note that the power $\Delta\gamma_{\pi/2}/\nu$ is determined by the exponents of a polymer near an infinite line (where $\alpha=\pi/2$), and is \emph{independent of the opening angle of the wedge} to which the polymer is attached. This behavior also exists in ideal polymers, where, in $d=2$ the angular function from Eq.~\eqref{eq:idealDistribution}, $\Theta_{\alpha}(\theta)=\cos(\theta\pi/2\alpha)$ \cite{Hammer2014,Alfasi2015}. Eq.~\eqref{eq:CardyAngular} was found to be in good agreement with extrapolation of enumeration of short SAWs, although small systematical discrepancies were observed \cite{Cardy1984}. 

\begin{figure}
\includegraphics{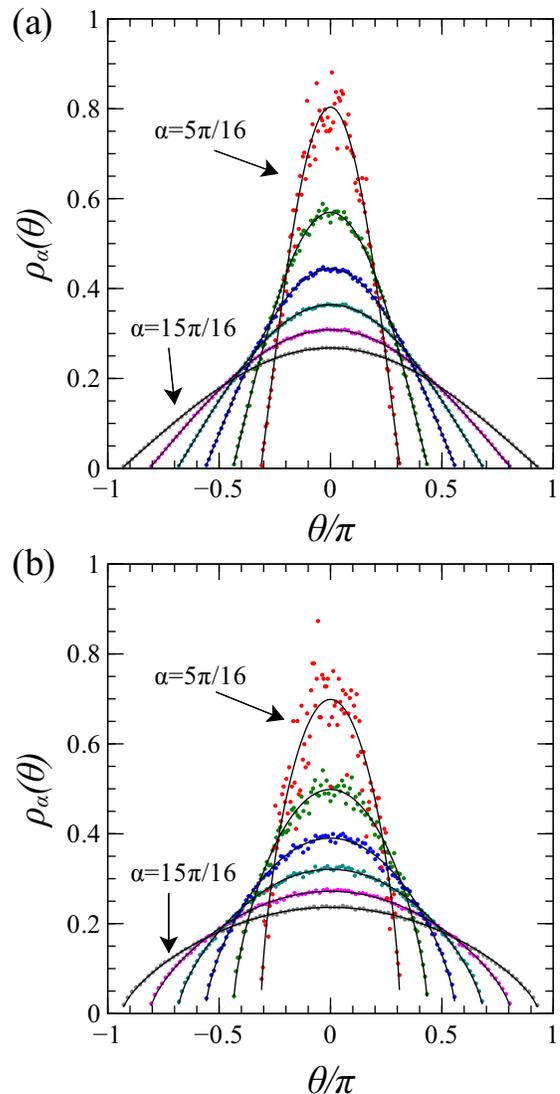}
\caption{(Color online) Angular probability distribution for the end-point of (a) a RW of size $N=10^6$ and (b) a SAW of size $N=5\times10^5$ starting near the corner of a two-dimensional wedge with opening angles $\alpha=j\pi/16$ where $j=5,7,..,15$. The continuous lines represent fits made to the functional form in Eq.~\eqref{eq:angDensFunction}.}
\label{fig:wedgeDens}
\end{figure}

\begin{figure}
\includegraphics{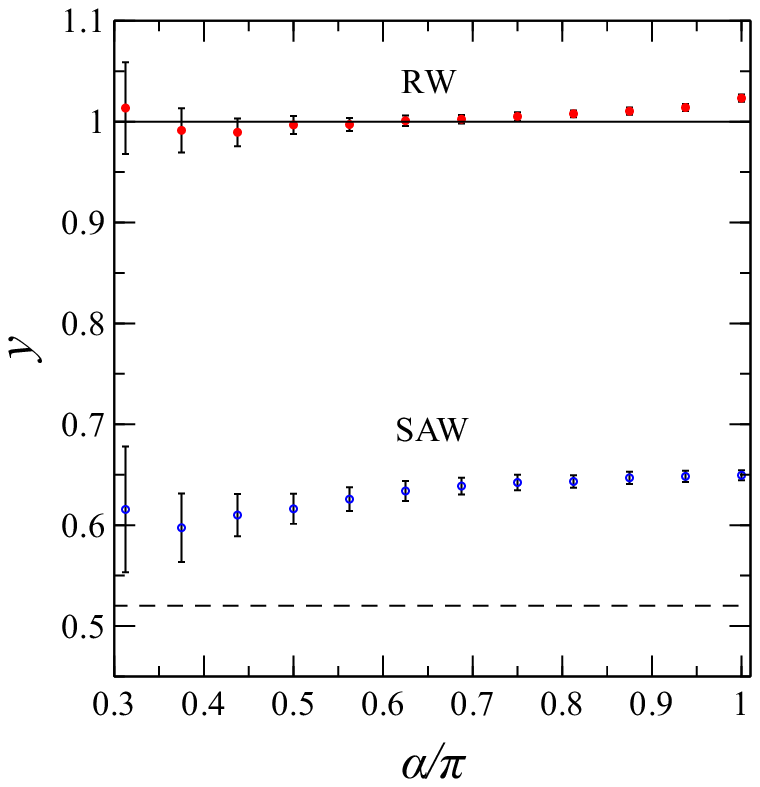}
\caption{(Color online) The power $y$ extracted from the fits in Fig.~\ref{fig:wedgeDens}, as a function of the opening angle of the wedge. The solid line denotes the known value $y=1$ for RWs. The dashed line denotes the prediction in \cite{Cardy1984} for SAWs.}
\label{fig:wedgePower}
\end{figure}

The angular density was measured for RWs with $N=10^6$ steps and SAWs with $N=5\times10^5$. In order to test the validity of Eq.~\eqref{eq:CardyAngular}, we attempted to fit the angular density to the form
\begin{equation} \label{eq:angDensFunction}
\rho_{\alpha}(\theta) = A\left[\cos\left(\frac{\theta\pi}{2\alpha}\right)\right]^y, 
\end{equation}
where $A$ and $y$ are the free parameters. In Fig.~\ref{fig:wedgeDens} we present $\rho_{\alpha}(\theta)$ along with the corresponding fits for selected opening angles $\alpha$. The power $y$ extracted from the fits is presented in Fig.~\ref{fig:wedgePower}. 
For RWs, we find that the power is close to the known value for infinite walks, $y=1$, with deviations that result from the fact that the walks in the simulation are finite. The functional form in Eq.~\eqref{eq:idealDistribution} is expected to hold when $h$, the distance between the starting point of the polymer and the corner of the wedge (ten lattice units in this case), is much smaller than the characteristic size of the polymer (say $R$). As can be seen in Fig.~\ref{fig:wedgePower}, for larger values of $\alpha$, where the polymer is less confined and $R$ is smaller, the deviation from the line $x=1$ is more significant. Note that for the SAWs in the simulation the characteristic sizes are much greater (since $\nu$ is greater) and we do not expect such corrections to be important.

\begin{figure}
\includegraphics[width=9cm]{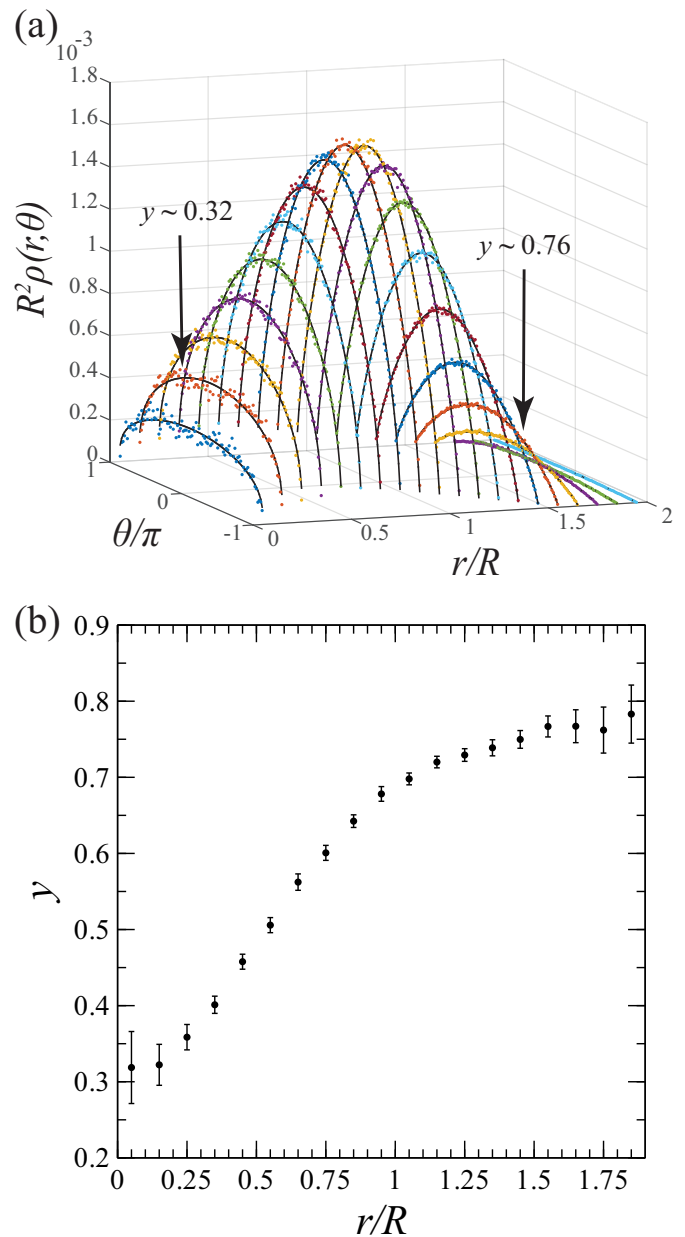}
\caption{(Color online) (a) End-point distribution function $\rho(r,\theta)$ of a SAW of $5\times10^5$ steps near an excluded semi-infinite line,  for various values of $r/R$, where $R^2$ is the mean square end-to-end distance of a SAW in free space (without the excluded line). The solid lines represents fits to the functional form in Eq.~\eqref{eq:angDensFunction}. (b) The power $y$ from the fits in (a).}
\label{fig:detailedEndPointDist}
\end{figure}

For SAWs, the power $y$ does not match the predicted value, $y=0.52$ \cite{Cardy1984}, and exhibits a weak dependence on the opening angle of the wedge. The small deviation is in fact consistent with the systematical errors reported in \citep{Cardy1984}.  The prediction in Eq.~\eqref{eq:CardyAngular} is based on the assumption that the dominant contribution for the integral in Eq.~\eqref{eq:angDistDef} comes from a region where $\rho_{\alpha}(r,\theta)$ can be separated into a product of angular and radial functions and the angular part has the form given in Eq.~\eqref{eq:CardyAngular}. This kind of separation is expected to occur for long polymers in the limits $r\gg R$ or $R\gg r \gg h$, since in these regions the system has no typical length scale. For long RWs, we know the separation into a product of radial and angular functions occurs for any $r$ (see Eq.~\eqref{eq:idealDistribution}). We measured the detailed end-point probability distribution $\rho(r,\theta)$ for a two-dimensional SAW near an excluded semi-infinite line. As can be seen in Fig.~\ref{fig:detailedEndPointDist}, the distribution \emph{does not} separate to a product of a radial and angular functions.
To get a sense of the angular dependence of $\rho$, we fitted the angular distribution at different radii to the form in Eq.~\eqref{eq:angDensFunction} and observed the changes in the power $y$ computed from the fits (Fig.~\ref{fig:detailedEndPointDist}b). For small $r/R$, $\rho$ is flatter, i.e., displays weaker variation with respect to $\theta$. For large $r/R$, the power $y$ saturates as expected. We find that $\rho_{\alpha=\pi}(r,\theta)$ can be roughly approximated by the function
\begin{multline} \label{eq:2PowerDist}
\rho_{\alpha}(r,\theta)\propto \\
\left\{x^{a_1}\left[\cos\left(\frac{\theta\pi}{2\alpha}\right)\right]^{b_1}+
Ax^{a_2}\left[\cos\left(\frac{\theta\pi}{2\alpha}\right)\right]^{b_2}\right\}e^{-Bx^4},
\end{multline}
where $x=r/R$ as before, $a_1=1.9$, $a_2=3.53$, $A=3.84$, $b_1=0.3$, $b_2=0.8$ and $B=0.44$. The power in the exponent in Eq.~\eqref{eq:2PowerDist} was assumed to be $\delta=1/(1-\nu)=4$, as is known for a SAW in free space and in accordance with the results of Sec.~
\ref{subsec:radialDist}. These values should not be taken too seriously. The function in Eq.~\eqref{eq:2PowerDist} is an example of a simple form of $\rho_{\alpha}(r,\theta)$ that separates into a product of angular and radial functions for $x\ll1$ and $x\gg1$ but displays non-trivial angular dependence for arbitrary $x$. The density in Eq.~\eqref{eq:2PowerDist} crosses over between two powers of the cosine at a typical radius that depends on $A$, $B$, $a_1$ and $a_2$. It seems that to properly describe $\rho_{\alpha}(r,\theta)$ for SAWs in wedges, more terms similar to the two in the curly brackets of Eq.~\eqref{eq:2PowerDist} with different powers and prefactors must be used.  

\section{Universal size ratio $R_{\alpha}^2/R^2$}
\label{sec:ratio}

The ratio between the mean square end-to-end distance $R_{\alpha}^2$ of the polymer attached to a cone/wedge and $R^2$ for a polymer in free space is a universal property which becomes independent of $N$ for large polymers. It was studied for a polymer near a flat surface both numerically \cite{Lax1974,Clark1978} and analytically \cite{Freed1983,Nemirovsky1985,Douglas1986,Douglas1987,Douglas1989}. Freed \cite{Freed1983} extended the renormalization group method to include polymers attached to a wall. The wall and the steric repulsion between monomers were introduced as perturbations to the ideal polymer model in free space. If the surface has dimension $d_{\parallel}=2$, both the steric repulsion and the boundary become irrelevant when the dimension of the system $d>4$ (see Sec.~\ref{sec:deltaGamma}). The end-point distribution function and the polymer size can then be evaluated by expanding in $\epsilon=4-d$. Douglas and Kosmas \cite{Douglas1989} treated a polymer near a surface of arbitrary dimension by introducing a second parameter $\epsilon_{\perp}=(2+d_{\parallel})-d$. They showed that to first order in $\epsilon$ and $\epsilon_{\perp}$ the universal ratio depends only on the dimensions of the system and the surface and on the exponent $\nu$. For a real polymer in $d=3$ near a plane, which corresponds to $\alpha=\pi/2$ in our description, they found $R_{\alpha}^2/R^2\approx7/6$. We measured a value of 1.22, in agreement with \cite{Douglas1986}.

\begin{figure}
\includegraphics{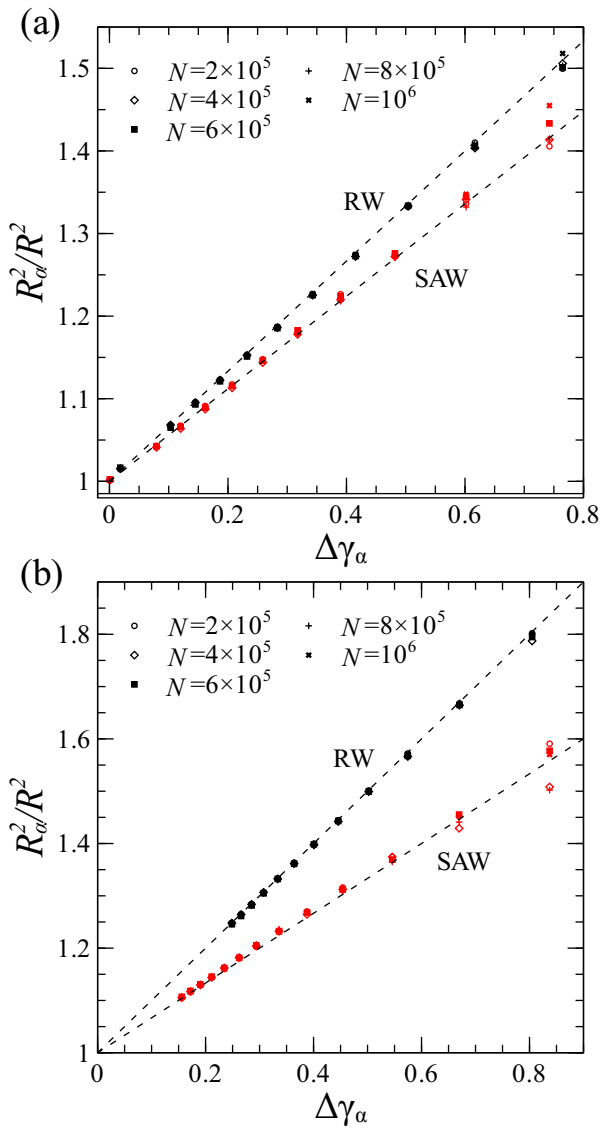}
\caption{(Color online) Mean square end-to-end distance $R^2_{\alpha}$ for RWs (black symbols) and SAWs (red symbols) of 5 sizes from $N=2\times10^5$ to $N=10^6$, starting from the tip of a repulsive cone in $d=3$ (a) or wedge in $d=2$ (b) with opening angle $\alpha$, divided by the corresponding mean square end-to-end distance of the walks in free space $R^2$. The ratio is shown as a function of the difference in the critical exponent $\gamma$ for a polymer in free space and a polymer attached to a cone/wedge. The dashed lines denote the known theoretical result for infinite RWs (Eq.~\eqref{eq:universalRatio}).}
\label{fig:Rsqr}
\end{figure}

For ideal polymers (RWs), Eq.~\eqref{eq:universalRatio} determines $R_{\alpha}^2/R^2$ and the $\alpha$ dependence of the ratio can be written in terms of $\Delta\gamma_{\alpha}$. 
Even though $\rho_{\alpha}(r)$ for SAWs is different from that which is known for RWs (see Sec.~\ref{subsec:radialDist}), we compared the universal ratio $R^2_{\alpha}/R^2$ measured for SAWs to the measured $\Delta\gamma_{\alpha}$. As is shown in Fig.~\ref{fig:Rsqr}, we find that Eq.~\eqref{eq:universalRatio} describes the relationship between these properties quite well, especially in $d=3$.
It is not surprising that the ratio $R_{\alpha}^2/R^2$ can be expressed in terms $\Delta\gamma_{\alpha}$, since both $R_{\alpha}^2/R^2$ and $\Delta\gamma_{\alpha}$ are monotonic functions of $\alpha$. However, it is notable that $R_{\alpha}^2/R^2$ is, to a good approximation, linear in $\Delta\gamma_{\alpha}$ for a wide range of opening angles, with the \emph{same} slope in terms of universal exponents as is known analytically for RWs. Quantitatively, we observe that 
\begin{equation} \label{eq:expansionInGamma}
R_{\alpha}^2/R^2=1+A_1\Delta\gamma+A_2\Delta\gamma^2+...,
\end{equation} 
where $A_1=1/d\nu$, while $A_2<0.01$ in $d=3$ and $A_2\sim0.1$ in $d=2$. We do not have an analytical basis for this result. Possibly, such a relation may emerge from higher order terms in the renormalization group approach.

\section{Summary and conclusions}
\label{sec:conclusions}
The generation of large ensembles of long chains has always presented a challenge in polymer simulations \cite{Milchev1998,Hsu2004}. This is especially true when attempting to measure universal properties, since the critical point in the polymer system is reached only in the limit $N\rightarrow\infty$. The recent implementation of the pivot algorithm by Clisby \cite{Clisby2010,Clisby2010b}, allowed a significant increase in polymer sizes feasible in simulations. 

In this work, we demonstrated that by using intersection tests with surfaces that take advantage of the SAW tree data structure (which lies at the heart of Clisby's implementation), it is possible to perform a direct study of large polymers near surfaces and measure their universal properties. We expect that this advancement can be used not only in the study of polymers in various geometries, but also in simulating polymers with different topology (e.g. star polymers \cite{Daoud1982} and polymer brushes \cite{DeGennes1987}). 

Scale free boundaries such as infinite wedges and cones constitute a special class of systems, where the universal properties change continuously (e.g. with the opening angle of the cone/wedge). This quality makes it possible to study study universal properties which are shape dependant like $\gamma_{\alpha}$ and $R_{\alpha}^2/R^2$ and find relations like that in Eq.~\eqref{eq:expansionInGamma}.

Our study of the end-point distribution of SAWs in wedges and cones revealed similarities in the radial behavior of the distribution $\rho_{\alpha}(r)$ to the ideal polymer case (especially regarding the ratio $R_{\alpha}^2/R^2$), but also an important difference in the angular dependence. The end-point distribution $\rho_{\alpha}(r,\theta)$ for SAWs is more complicated than that of RWs and generally does not separate into a product of angular and radial functions.

\begin{acknowledgments}
We thank M. Kardar for useful discussions and for comments on the manuscript. We thank the referees for the detailed and constructive remarks. This work was supported by the Israel Science Foundation under grant 186/13.
\end{acknowledgments}

\bibliographystyle{apsrev4-1}

\end{document}